# Accelerating Neural Networks for Large Language Models and Graph Processing with Silicon Photonics

Salma Afifi, Febin Sunny, Mahdi Nikdast, Sudeep Pasricha
*Department of Electrical and Computer Engineering*
Colorado State University
Fort Collins, Colorado, USA
{salma.afifi, febin.sunny, mahdi.nikdast, sudeep}@colostate.edu

*Abstract*—In the rapidly evolving landscape of artificial intelligence, large language models (LLMs) and graph processing have emerged as transformative technologies for natural language processing (NLP), computer vision, and graph-structured data applications. However, the complex structures of these models pose challenges for acceleration on conventional electronic platforms. In this paper, we describe novel hardware accelerators based on silicon photonics to accelerate transformer neural networks that are used in LLMs and graph neural networks for graph data processing. Our analysis demonstrates that both hardware accelerators achieve at least 10.2× throughput improvement and 3.8× better energy efficiency over multiple state-of-the-art electronic hardware accelerators designed for LLMs and graph processing.

## I. INTRODUCTION

Large language models (LLMs) have garnered significant attention in recent years because of their exceptional performance in natural language processing (NLP) tasks. LLMs are based on the Transformer neural network architecture which attributes its capabilities to the self-attention mechanism [1]. Transformer networks exhibit versatile capabilities in excelling at and addressing a variety of tasks, such as transduction and computer vision, leading to a growing adoption across various fields. This is unlike earlier models restricted to specific task domains such as recurrent neural networks (RNNs) and convolution neural networks (CNNs). Therefore, LLMs are widely expected to lead us closer to the artificial general intelligence paradigm.

In addition, the success of deep learning techniques has led to applying neural network architectures to graph processing. However, conventional deep neural networks (DNNs) are designed to operate on regular grid-like patterns within Euclidean space, complicating their deployment with non-Euclidean data domains such as graphs. Thus, graph neural networks (GNNs) have emerged as a revolutionary approach for graph processing, attaining remarkable performance in many tasks such as social network relationship analysis and knowledge graphing [2]. Despite their complex structure, GNNs are crucial to enable the development of cutting-edge AI applications and systems.

The development of hardware platforms capable of providing sufficient support to execute transformers in LLMs and GNNs for graph processing with high performance, while abiding by strict power constraints, is thus of paramount importance. Although hardware acceleration for neural networks such as CNNs and RNNs has been extensively studied, the processing of transformer and GNN models presents unique challenges due to their greater complexity.

This research was supported by the National Science Foundation (NSF) under grants CNS-2046226 and CCF-1813370

Moreover, conventional electronic accelerators are increasingly susceptible to the limits of the post Moore's law era, where diminishing performance gains are evident as technology scales [3]. These limitations pose significant performance and energy efficiency obstacles when executing LLMs and GNNs on electronic hardware platforms.

Silicon photonics has demonstrated its proficiency not only in high-throughput communication within the telecom and datacom sectors but also as a viable solution for chip-scale communication. The incorporation of energy-efficient optical modulators enables photonic interconnects to achieve elevated modulation speeds (>50 GHz) and maintain low energy consumption (<70 fJ/bit) [4]. Moreover, CMOS-compatible silicon photonic components can be used for computations, such as matrix-vector multiplications and logic gate implementations [5]. Optical computations show the potential to achieve $O(N)$ energy scaling for $O(N^2)$ fixed-point operations while energy tends to increase with $O(N^2)$ for digital electronic systems [6].

In this paper, based on our recent works [2], [7], we present the first hardware accelerators for neural networks used in LLMs and graph processing that leverage silicon photonics. Unlike other accelerators, our architectures are carefully tailored for a wide range of building block components leveraged in LLMs and graph processing. By integrating the high-speed, low-energy optical operations and harnessing the inherent parallelism of silicon photonics, our proposed accelerators demonstrate significant enhancements in both throughput and energy efficiency when compared to the state-of-the-art LLM and GNN hardware accelerators.

## II. LARGE LANGUAGE MODELS

LLMs make use of Transformer neural network models characterized by massive parameter sizes and outstanding learning capabilities. The original transformer model [8] forms the architecture backbone for many state-of-the-art LLMs. Fig. 1 shows the transformer model composed of two main blocks: an encoder and a decoder. The encoder transforms the input sequence into a continuous abstract representation. Subsequently, the decoder processes this representation incrementally, generating a singular output while incorporating prior outputs.

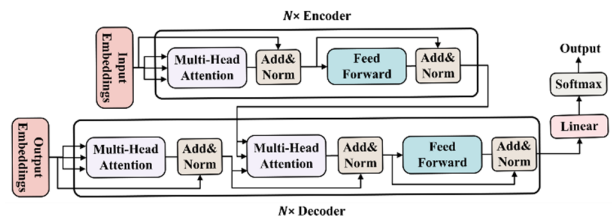

Fig. 1: Transformer neural network architecture overview.

As shown in Fig. 1, the encoder and decoder blocks are usually composed of $N$ stacked layers and their main sub-blocks are the multi-head attention (MHA) and feed forward (FF) layers, along with residual connections for each and followed by layer normalization. The computation of the most intricate operation within the transformer, namely the self-attention mechanism, is carried out within the MHA block. Each MHA has $H$ self-attention heads, and each attention head generates the query ($Q$), key ($K$), and value ($V$) vectors to compute a scaled dot-product attention as follows:

$$Head(X) = attention(Q,K,V) = softmax\left(QK^T/\sqrt{d_K}\right)V, \quad (1)$$

where $X$ is the input matrix and $d_K$ is the dimension of $Q$ and $K$. The output of the MHA is the concatenation of the self-attention heads' outputs, followed by a linear layer. The FF network is composed of two dense layers with a *RELU* activation in between. Pre-trained language models based on the transformer architecture typically either include just the encoder block, such as in BERT [9], or just the decoder block, such as in GPT [1]. Additionally, the vision transformer (ViT) models used in computer vision tasks have $N$ encoder layers followed by a multi-layer perceptron [2]. Such variations in Transformers across NLP and vision tasks along with the complex operations and structure of Transformers impose serious challenges for their acceleration.

To mitigate memory-wall bottlenecks in Transformers, specifically those used in LLMs, prior works have explored processing-in-memory (PIM). An in-memory computing-based transformer accelerator called TransPIM was presented in [10] along with a novel dataflow for optimized data movements to high bandwidth memory. The work in [11] proposed a ReRAM-based PIM accelerator for transformers. However, PIM introduces a distinct set of challenges. Specifically, ReRAM cells encounter various reliability issues related to endurance and retention [12]. Other works presented FPGA-based solutions as in [13] where a hardware accelerator was designed to enhance the performance of MHA and FF layers. Their strategy involves partitioning the weight matrices utilized in both MHA and FF layers to enable resource sharing between the layers. An FPGA-based acceleration framework was also presented in [14].

## III. GRAPH PROCESSING

By incorporating deep learning into the domain of graph processing, GNNs have substantially transformed numerous tasks and applications relying on graph structures, such as node classification, link prediction, and graph classification. GNNs leverage the connections inherent in a graph to comprehend and represent the relationships among vertices. They employ an iterative methodology based on the graph's structure and incorporate edges, vertices, and graph feature vectors, representing the known attributes of these elements.

Fig. 2 illustrates the three main stages involved in a GNN processing. First, a graph is input to the GNN that is usually preprocessed offline for purposes such as sampling the graph. Second, the aggregation stage iteratively collects the neighbors of each vertex and subsequently reduces all the gathered data into a singular vector. The data can be reduced by using various arithmetic functions such as summation, mean, or maximum. Lastly, the resultant feature vectors are passed through the combination phase which is usually composed of a neural network.

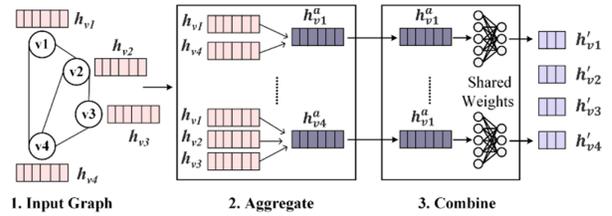

Fig. 2: An overview of GNN inference: 1) Input graph showing initial feature vectors; 2) Aggregation phase, where each node's neighbors are reduced to one feature vector; 3) Combine and Update phases, where each node is linearly transformed and updated using a non-linear activation function.

Accelerating GNNs involves overcoming several formidable challenges. While most GNNs follow the same stages outlined earlier, various new and more complex GNN algorithms and models have recently been introduced. For example, graph convolution networks (GCNs) [15] extend convolution to the graph space by having it defined on irregular graph structures. Various models based on GCNs have also emerged such as GraphSAGE and graph isomorphism network (GIN). Graph attention networks (GATs) represent another category of GNNs where the models integrate an attention mechanism and enhance node features by employing a pairwise function between nodes, integrating learnable weights [2]. A system accelerating GNNs needs to have the capability to accommodate such diversity found in GNN models. Furthermore, GNNs present a set of both dense and very sparse computations and real-world graphs can be extremely large and highly irregular. Accordingly, GNNs frequently necessitate extensive memory bandwidth and multiple irregular memory accesses.

Several hardware accelerators for GNNs have been developed in recent years. An electronic accelerator was proposed in [16] where the core unit of their architecture consists of an aggregator module, a DNN accelerator, a DNN queue, and a graph processing element. EnGN [17] utilizes clustered processing elements to handle GNNs in a single dataflow as concatenated matrix multiplication of feature vectors, adjacency matrices, and weights. Other accelerators such as HyGCN [18] and GRIP [19] are composed of separate engines for aggregation and combination. Multiple ReRAM and PIM accelerators have also been presented such as ReGNN [20] and ReGraphX [21].

## IV. SILICON PHOTONICS FOR NEURAL NETWORK ACCELERATION

Neural network accelerators built with silicon photonics have emerged as a powerful alternative to electronic-based accelerators due to their significant energy and performance improvements [3]. Accordingly, the utilization of silicon photonics for enhancing the acceleration of neural networks has garnered considerable attention [22]-[27].

CrossLight [28] is a CNN optical accelerator that benefits from cross-layer device-level, circuit-level, and architecture-level optimization. The CrossLight architecture is composed of dedicated vector-dot-product units for convolution and fully connected layers. In [29], an optical accelerator for sparse neural networks was introduced with a modular and a vector granularity-aware structure that enables significant throughput and energy improvements. Additionally, an optical hardware accelerator for RNNs was proposed in [30]. Several other architectures have focused on optical-domain acceleration for CNNs, RNNs, and MLPs [26]. To the best of our knowledge, the accelerators we proposed in [2] and [7] are the first silicon photonics-based accelerators for GNNs and transformers, respectively.

Neural network acceleration using silicon photonics requires imprinting the network's parameters onto optical signals which enables performing MAC operations. This can be done using coherent or non-coherent implementations. Coherent architectures utilize a single wavelength where the parameters are imprinted onto the optical signal's phase. On the other hand, multiple wavelengths are leveraged in non-coherent architectures and the parameters are imprinted onto the optical signal's amplitude. Our optical accelerators' core operations are performed using opto-electronic microring resonator (MR) devices. Each MR can be designed and tuned to work at a specific wavelength, called MR resonant wavelength ($\lambda_{MR}$), defined as:

$$\lambda_{MR} = \frac{2\pi R}{m} n_{eff}, \qquad (2)$$

where $R$ is the MR radius, $m$ is the order of the resonance, and $n_{eff}$ is the effective index of the device.

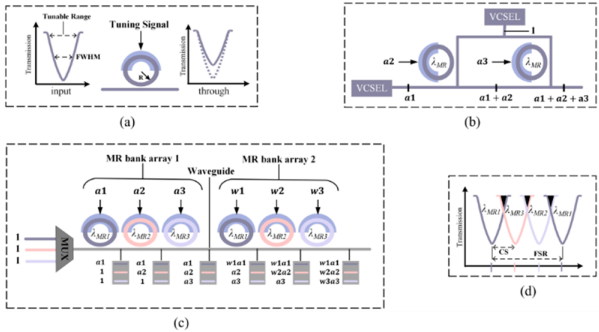

Fig. 3: (a) MR input and through ports' wavelengths after imprinting a parameter onto the signal; (b) two MR devices used to perform optical coherent summation to add values a$_1$, a$_2$, and a$_3$; (c) MR bank arrays used to perform multiplication by imprinting input vector (a$_1$-a$_3$), followed by weight vector (w$_1$-w$_3$); (d) MR bank response and heterodyne crosstalk shown in black, where CS is channel spacing and FSR is free spectral range. From [2].

A tuning circuit can be used to carefully alter $n_{eff}$ which modulates electronic data onto the optical signals. The tuning circuit should result in a change in $n_{eff}$, and thus a resonant shift ($\Delta\lambda_{MR}$). Fig. 3(a) illustrates the transmission plots for the input and the through ports' responses after a parameter is imprinted onto the input signal. As a result of tuning an MR's $\lambda_{MR}$, a predictable change in the optical signal's amplitude can be observed. Our accelerators expand upon this to implement their two main operations: summation and multiplication, as discussed below.

Fig. 3(b) shows an example of performing the addition of $a1$, $a2$, and $a3$ using coherent summation where a single wavelength $\lambda_{MR}$ is used. All MR devices are configured to operate on the same resonant wavelength. VCSEL units are laser sources that can be configured to generate an optical signal with a certain wavelength and an amplitude specified by an input analog signal. When the three optical signals meet, they undergo interference, resulting in a summation operation and the final output $a1 + a2 + a3$ is computed.

An example of performing multiplication is shown in Fig. 3(c). To enhance throughput and emulate neurons, non-coherent silicon photonics is used where multiple optical signals with different wavelengths are multiplexed into the same waveguide using wavelength division multiplexing (WDM). This waveguide then passes through two banks of MR devices where each MR in each bank is configured to operate on a specific wavelength (outlined by the different colors). In the example shown, three different wavelengths are needed to multiply the input activation vector $[a1, a2, a3]$ by the weight vector $[w1, w2, w3]$. The first set of the MR devices imprint the input activation values onto the three different wavelengths. When the second set of the MR devices imprint the weight values onto the same optical signals, a multiplication operation occurs, and the output optical signals represent the output vector $[w1a1, w2a2, w3a3]$.

Non-coherent silicon photonics incurs various challenges such as high energy overheads and heterodyne (or incoherent) crosstalk. The latter is shown as the shaded regions in black in Fig. 3(d). The crosstalk arises when a segment of an optical signal from adjacent wavelengths interferes with another wavelength. We thoroughly discuss our efforts to overcome such challenges in Section V.B.

V. TRANSFORMER AND GNN HARDWARE ACCELERATORS

Accelerating Transformers and GNNs requires having an efficient underlying hardware that is capable of accommodating and tailoring its execution to each class of the neural networks' unique requirements. In this section, we present our silicon-photonic-based accelerators for Transformers and GNNs. Section V.A discusses the tuning circuit design used in our accelerators, followed by Section V.B which introduces the various MR device optimizations performed. Sections V.C and V.D then present our Transformer and GNN optical hardware architecture designs.

*A. Tuning Ciruit Design*

As discussed in section IV, MR devices require a tuning circuit that can be based on the electro-optic (EO) or thermos-optic (TO) effect. EO tuning operates at a faster rate and consumes less power, but cannot be used for large tuning ranges. On the other hand, TO tuning accommodates larger tunability range but has the drawback of higher latency and power consumption [30]. We have employed a hybrid tuning approach for our hardware accelerators, merging the benefits of EO and TO while mitigating their respective drawbacks. In our approach, EO tuning is leveraged for fast induction of small $\Delta\lambda_{MR}$, whereas slower TO tuning is only enabled infrequently when there is a need for larger $\Delta\lambda_{MR}$. Additionally, our designs integrate the thermal eigenmode decomposition method (TED) as outlined in [29], to effectively decrease the power consumption associated with TO tuning and mitigate thermal crosstalk.

*B. MR Device Optimization*

Operating in the analog photonic domain presents a set of noise sources that need to be addressed to ensure correct execution of Transformers and GNNs. Such noise sources include thermal crosstalk, heterodyne (or incoherent) crosstalk, and homodyne (or coherent) crosstalk. Our TED-based method (Section V.A) mitigates the thermal crosstalk between TO tuning circuits. Heterodyne or inter-channel crosstalk emerges when multiple wavelengths are used in the same waveguide employing non-coherent silicon photonics. This occurs when a portion of an optical signal from a neighboring wavelength undesirably leaks into the MR spectrum of another wavelength, causing inaccuracies. To efficiently alleviate heterodyne crosstalk, spectral overlap needs to be minimized. To achieve this, various factors should be considered and optimized. Specifically, key elements such as well-designed channel spacing, Q-factor tuning, ensuring a signal-to-noise ratio (SNR) in the output that surpasses photodetector sensitivity, and optimizing the tunable range of the designed MRs must be addressed.

On the other hand, homodyne crosstalk occurs due to the presence of undesired coupling between signals with the same wavelength. As will be discussed later, some of our accelerators' computation circuits are based on coherent silicon photonics. A portion of a signal might end up leaking into an opto-electric device and experiences a change in its phase. These leaked signals then interfere with the output signals causing inaccuracies and impacting the SNR. One way to mitigate homodyne noise is to increase the crossover coupling by increasing the gap between the MR input waveguide and MR ring waveguide. This reduces the amount of crosstalk signal being coupled over from the MR to the main waveguide. For achieving this, several MR configuration parameters need to be carefully tuned.

As thoroughly explained in [2] and [7], we accurately modeled heterodyne crosstalk, homodyne crosstalk, and the MR's device configuration. Utilizing these models and the simulation tool suite from Ansys Lumerical [31], we can identify the design space for our MRs and the MR banks they constitute. Accordingly, for our accelerators, we have determined the optimal MR design and configurations that would result in negligible crosstalk noise.

### C. Transformer Hardware Accelerator Design

Our Transformer hardware accelerator is a photonic accelerator capable of accelerating the inference of a broad family of neural networks used in LLMs. Fig. 4 presents an overview of the photonic accelerator's architecture, with MHA and FF units being the core components.

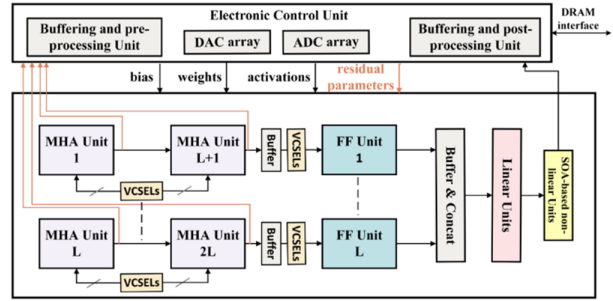

Fig. 4: Transformer neural network architecture overview. From [7].

The MHA unit enables performing optically the time-consuming matrix multiplications (MatMuls) which constitutes the major challenge with most LLMs' inference. To minimize intermediate storage and opto-electronic conversions, our architecture decomposes the main MatMul in MHA (see (1) in Section II) as follows:

$$Q.K^T = Q.(X.W_K)^T = (Q.W_K^T).X^T. \qquad (3)$$

Such decomposition mitigates the need to convert the optical signals (matrix $K$) to the digital domain to perform its transpose operation before the multiplication with matrix $Q$. Conversely as outlined in Fig. 5(a), matrices $X$, $W_Q$, $W_K^T/\sqrt{d_K}$, and $X^T$ are computed and stored offline, which allows us to perform the MatMul completely in the optical domain. After computing $Q.K^T$ with the upper MR bank arrays illustrated in Fig. 5(a), the accumulated partial sums are processed using balanced photodetectors (BPDs). BPDs facilitate the handling of both positive and negative parameter values by incorporating distinct positive and negative arms within the same waveguide. The sum obtained from the negative arm is subtracted from the sum originating from the positive arm. Subsequently, the results are converted to the digital domain to undergo softmax computation using lookup tables (LUTs) and simple digital circuits. The linear layer is implemented optically using two MR bank arrays, while the residual connections are executed by adding the MHA input to its current output using coherent photonic summation. Lastly, layer normalization (LN) is implemented optically using a single MR, tuned by the LN parameter. The entire MHA architecture is shown in Fig. 5(b).

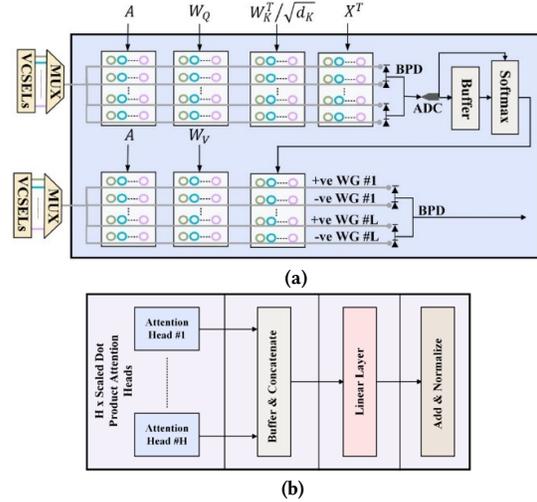

Fig. 5: (a) Attention head unit comprised of seven MR bank arrays for MatMul operations, each with dimension K×N; (b) MHA unit composed of H attention heads, buffer and concatenate block, linear layer, and an add and normalize block. From [7].

### D. GNN Hardware Accelerator Design

The architecture of our GNN accelerator is depicted in Fig. 6. Its primary components (aggregate, combine, and update) are partitioned into $V$ execution lanes. During the inference phase, each lane is tasked with processing one output vertex concurrently with all other lanes. The aggregate block collects all neighboring vertices and their corresponding edge data and executes a reduce function for each assigned output vertex. The combine block then applies a linear transformation to each aggregated vertex feature vector. Lastly, the update block employs a non-linear activation function to derive the updated vertex feature vectors $h_v'$.

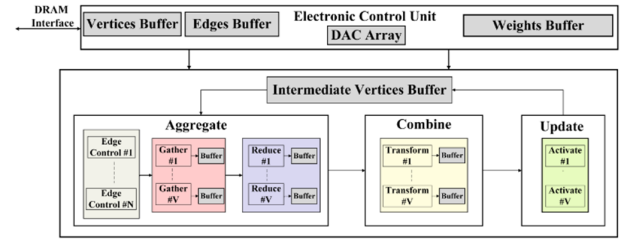

Fig. 6: Overview of optical GNN accelerator architecture. From [2].

One key performance challenge when accelerating GNNs is handling the highly sparse and irregular memory accesses. By employing a "buffer and partition" optimization as described in [2], we can alleviate this bottleneck. This technique dictates splitting the input graph into blocks of $N$ and $V$ where the aggregate block then is composed of $N$ edge control units, $V$ gather units, and $V$ reduce units. Each execution lane is assigned one output node per cycle while $N$ input nodes are fetched by the edge control units and forwarded to the gather units as needed. Gather units then convert this data (output node and neighboring input nodes) to analog signals that are used to tune the MRs.

To accommodate a wide range of reduction functions (summation, mean, and maximum), the reduce unit is configured as an optical coherent summation block (see Fig. 7(a)). Each row in the reduce unit corresponds to a feature from the vertex feature vectors while each column is associated with a neighbor input vertex. The signals produced by the top VCSELs traverse the MR bank to imprint the feature values of neighboring nodes onto the signals. As discussed in Section IV, when the different waveguides carrying signals with the same wavelength meet, they undergo interference. Accordingly, the result of each row is the summation of all the neighbor node values imprinted onto each optical signal. The combine block gathers outputs from the aggregate block and executes a linear transformation using learned weight parameters. This linear transformation within the transform unit occurs in the optical domain using MR bank arrays, as depicted in Fig. 7(b).

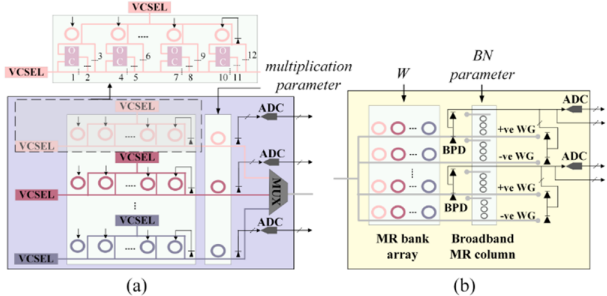

Fig. 7: (a) Reduce unit showing the needed changes in each feature lane to support the max aggregation operation using an optical comparator; (b) detailed view of transform unit. From [2].

Given that linear transformation operations in GNNs are predominantly matrix-vector multiplications, they can be accomplished in the optical domain using MR bank arrays. This involves passing the weight values to the transform unit as analog signals to fine-tune each MR while the feature vector values are imprinted onto the optical signals in the waveguide from the reduce units. Unlike the MR arrays utilized in reduce units for summation operations (see Fig. 7(a)), multiplications are performed non-coherently.

Lastly, the update block comprises $V$ update units, each tasked with applying a non-linear activation function to the output originating from the transform units. Non-linear activation functions such as *RELU*, *sigmoid*, and *tanh* are implemented optically using semiconductor-optical-amplifiers (SOAs). On the other hand, non-linear activation functions that pose difficulties to be performed optically (e.g., softmax), are implemented using LUTs and simple digital circuits. Additionally, the hardware accelerator supports various orchestration and scheduling optimization techniques including graph buffering and partitioning, execution pipelining and scheduling, weight DAC sharing, and workload balancing [2]. Such techniques enable the hardware to efficiently accommodate a broad family of GNN models and regulate the memory accesses during execution.

## VI. EXPERIMENTAL RESULTS

For evaluating our proposed hardware accelerators, we developed comprehensive simulators in Python to estimate the power and latency costs. The simulators efficiently perform software mapping for the various Transformer and GNN models used in our experiments, as well as hardware mapping where all the optoelectronic and electronic devices and circuits are modeled. For all the memories and buffers employed in our accelerators, CACTI [32] was used to obtain their performance and energy estimates. Multiple Transformer and GNN models and datasets were used in our experiments. Based on our analysis conducted for each model and dataset, we concluded that employing 8-bit model quantization yields algorithmic accuracy comparable to models utilizing full (32-bit) precision. Consequently, we focused on the acceleration of Transformer and GNN models with 8-bit precision. The specific architectural details of each hardware accelerator such as the numbers of the computational blocks, were determined through detailed design-space analysis as explained in [2] and [7].

Our proposed accelerators are compared against multiple computing platforms and state-of-the-art hardware accelerators. Our Transformer accelerator (referred to as *TRON*) is compared against Tesla V100-SXM2 GPU, TPU v2, Intel Xeon CPU, TransPIM [10], FPGA transformer accelerator in [13] (FPGA_Acc1), VAQF [33], and FPGA transformer accelerator in [14] (FPGA_Acc2). For our GNN accelerator (referred to as *GHOST*), the platforms chosen for comparison were: GRIP [19], HyGCN [18], EnGN [17], HW_ACC [16], ReGNN [20], ReGraphx[21], TPU v4, Intel Xeon CPU, and NVIDIA A100 GPU. We utilized reported power, latency, and energy values for the chosen accelerators, and directly acquired outcomes from model executions on the GPU, CPU, and TPU platforms to calculate the Energy Per Bit (EPB) and Giga Operations Per Second (GOPS) for each model and dataset.

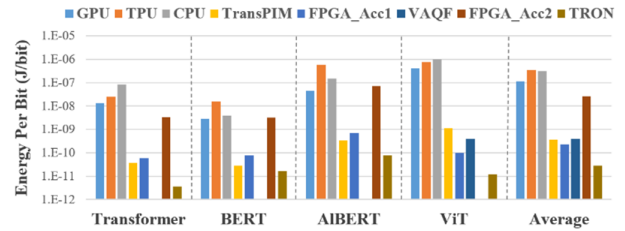

Fig. 8: EPB comparison across LLM accelerators [7].

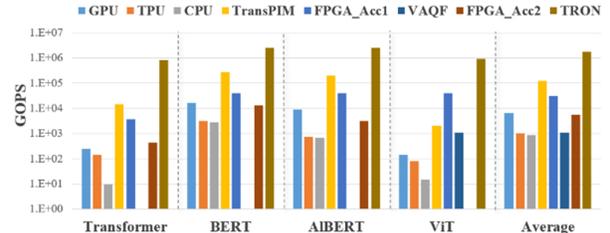

Fig. 9: Throughput comparison across LLM accelerators [7].

Figs. 8 and 9 show the EPB and GOPS comparisons between *TRON* and the other computing platforms and Transformer accelerators considered. On average, our architecture achieves at least 14× better throughput and 8× better energy efficiency. The increased throughput and EPB enhancements can be ascribed to *TRON's* fast execution in the optical domain and minimal conversions to the electronic domain.

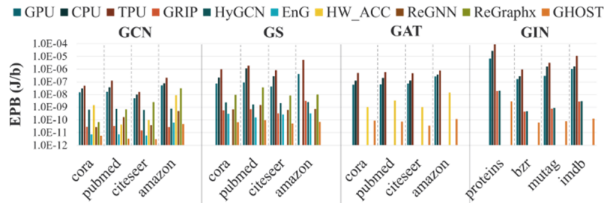
Fig. 10: EPB comparison across GNN accelerators [2].

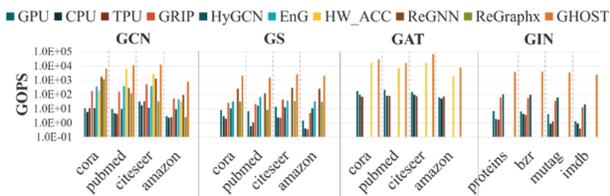
Fig. 11: Throughput comparison across GNN accelerators [2].

The EPB and GOPS comparisons for *GHOST* are shown in Figs. 10 and 11. Our simulation experiments reveal that *GHOST* outperforms existing hardware accelerators for GNNs, exhibiting a minimum of 10.2× improvement in throughput and 3.8× greater energy efficiency compared to GPU, TPU, CPU, and the various state-of-the-art GNN hardware accelerators considered. These findings highlight *GHOST's* proficiency in handling diverse graph processing tasks and its ability adapt its execution and memory accesses scheduling based on the specific GNN and dataset employed.

Overall, the improvements from our hardware accelerators can be attributed to the considerably low latency compute operations and data transfers in the optical domain, relatively low power consumption, and the various cross-layer optimization techniques employed.

## VII. CONCLUSION

In this paper, we presented two groundbreaking silicon photonic hardware accelerators for LLMs and graph processing. The design of these accelerators incorporates various device-level, circuit-level, and architecture-level optimizations. Our photonic hardware LLM accelerator exhibited at least 14× better throughput and 8× better energy efficiency compared to previously proposed Transformer accelerators. Our photonic graph processing accelerator showed a minimum of 10.2× throughput improvement and 3.8× better energy efficiency against state-of-the-art GNN accelerators. The achieved results underscore the use of silicon photonics for energy-efficient and high-throughput inference acceleration in LLMs and GNNs. Moving forward, both architectures pave the way for future research and advances in silicon photonic accelerators, with considerations for addressing open challenges such as alternative non-volatile optical memory cells, fabrication-process variations, and further optimizations at the device and circuit levels.